\documentstyle[12pt]{article}
\title{Unravelling quantum carpets: a travelling 
wave approach}
\author{Michael J. W. Hall\footnotemark[1] \footnotemark[2], 
Martina S. Reineker\footnotemark[1] and Wolfgang
P. Schleich\footnotemark[1]\\ \\ \footnotemark[1] Abteilung 
f\"{u}r Quantenphysik\\
Universit\"{a}t Ulm\\D-89069 Ulm, Germany\\ \\
\footnotemark[2] Theoretical Physics, IAS\\Australian National 
University\\ Canberra ACT 0200, Australia}
\begin{document}
\maketitle
\begin{abstract}
Generic channel and ridge structures are known to appear in the 
time-dependent position probability distribution of a one-dimensional 
quantum particle confined to a box.  These structures are shown to 
have a detailed 
quantitative explanation in terms of a travelling-wave 
decomposition of the probability density, wherein each 
contributing term corresponds simultaneously to 
(i) a real wave propagating at
a quantised velocity and (ii) to the time-averaged structure of 
the position distribution along a quantised direction in spacetime.  
The approach 
leads to new 
predictions of channel locations, widths and depths, and is able to
provide more 
structural 
details than earlier approaches based on partial interference and 
Wigner functions.
Results are also applicable to light diffracted by a periodic grating,
and to the quantum rigid rotator.
\end{abstract}
\newpage
\section{Introduction}

The position probability distribution $P(x,t)$ of a one-dimensional
quantum particle 
may be represented as a probability 
landscape 
in spacetime, where hills and valleys correspond to regions
of high and low probability density respectively. It has
recently been discovered that generic two-dimensional
structures appear across this probability landscape for a
wide variety of potentials and initial conditions
\cite{kinzel}-\cite{stift}. The highly-patterned 
nature of these
structures has led to them being called ``quantum carpets".

 For the simplest case, of a one-dimensional
particle moving freely between two end walls, the  carpet
patterns
are linear, with quantised slopes and intercepts
\cite{kinzel,berry,gross} 
More generally, however, the patterns
are curved for particles moving under the 
influence of a potential \cite{stift,kap,loinaz}.  Colour plots of 
quantum carpets, for a range of examples, may be found in \cite{stift} and
\cite{kap}.

Analogous carpet structures also arise for
light diffracted by a one-dimensional 
periodic grating in the paraxial approximation \cite{klein,boden}.
In this case the time variable $t$ is replaced by the propagation
distance $z$ of the diffracted beam, and $P(x,z)$ is the intensity
distribution of the beam.  Further, it will be seen that a generalised
form of these carpet structures arises for the angular distribution 
$P(\phi, t)$ of the two-dimensional quantum rigid
rotator.

Approaches based on interference between energy amplitudes
\cite{berry,gross}, Wigner functions \cite{stift}, and 
Greens function degeneracies \cite{kap,irena} have been 
used to explain the observed carpet patterns for various 
cases, and in particular the observed quantisation of slopes of the 
linear structures for the
confined one-dimensional particle is well understood.  However, 
even in this relatively simple case a number of generic features 
have not 
yet been provided with a general explanation, including the 
depths and widths of channel 
structures, the decrease of the latter two quantities with slope, 
and the ``chopped" nature of ridge structures (see Fig. 1).  

It will be shown in section 2 that such features can be understood to a 
limited extent by generalising an approach first developed by Berry 
\cite{berry}, based on the destructive interference of
energy amplitudes (see also \cite{gross}). This generalised {\it destructive 
interference} approach in particular 
predicts valley structures along the lines
\begin{equation} \label{berry}
x = k Vt + l L,  \hspace{1cm} kl = {\rm even\; integer}
\end{equation}
in spacetime for a particle confined to a one-dimensional box with 
end 
walls at $x=0$ and $x=L$,  providing that the energy amplitude 
differences 
$| \psi_n - \psi_{n+k}|$ of the wavefunction are sufficiently 
small 
for all $n$. Here $k$ and $l$ are integers, and 
\begin{equation} \label{vee}
V=\pi\hbar/(2ML)
\end{equation} 
is a natural speed defined via
the particle mass $M$. 

The destructive interference approach can be further used
to show that the average channel depth is expected to decrease 
as $k$ increases (important in resolving the Olbers-type
paradox that any channels at all can be 
discerned, given that an infinite number of them are predicted 
by (\ref{berry}) in any 
given spacetime region).  Moreover, as shown in the Appendix, it may be 
generalised to predict the locations of channel and ridge
structures for particles moving in {\it arbitrary} one-dimensional
potentials, with results in agreement with \cite{kap}. 

However, the above approach is {\it not} able 
to explain a number of other observed features.  
First, it is completely silent on the location of structures when the
condition of small energy amplitude differences is not met.  Second, as
will be shown by example, it fails to predict {\it all}
observed channel and ridge structures even when this condition {\it
is} met.  Finally, the approach yields no
information on the shapes and widths of the various 
structures. A new approach is thus clearly called for.

Such an approach is developed in section 3, based on a 
representation of the probability  landscape as a superposition of
travelling waves with velocities which are integer multiples of $V$,
propagating against a constant background.
The primary usefulness of this decomposition is that the {\it average}
of the probability distribution along any given direction in spacetime 
(eg, corresponding to a channel or ridge) is described by at most a  
{\it single} one of these travelling waves. In particular, the wave 
corresponding to velocity $kV$ 
provides specific, exact information on the average
locations, shapes, depths and widths of all channel and ridge structures with 
slope $kV$.  Connections with the destructive interference approach are
briefly discussed.

The averaging of $P(x,t)$ along particular spacetime directions, to predict
locations of corresponding channel structures, was first (independently)
suggested by
Berry and Bodenschatz \cite{boden} in the optical grating context, who
were also motivated by limitations of the destructive interference
approach.  That these averages arise naturally from a travelling wave
decomposition of $P(x,t)$, and can hence be simply superposed 
to recover the complete probability landscape (and to approximate
it with arbitrary accuracy by considering only waves up to a given
maximum speed) has not previously been realised. 

In section 4 it is shown further 
that each travelling wave can be written as the 
sum of two Wigner functions plus an
energy amplitude term.  This form is useful for showing how localisation
and symmetry properties of the initial probability distribution can
enhance or suppress carpet patterns.  In Sec.~5 predictions are verified
via the examples of a uniform initial wavefunction and an approximately
Gaussian initial wavefunction, for which analytic expressions are
obtained for the travelling waves.

Generalisations of the results to periodic optical gratings and to the quantum
rigid rotator are given in section 6, with conclusions presented in 
section 7.

\section{Destructive interference approach} 

\subsection{Derivation}

Berry derived the locations of channel 
structures of a one-dimensional particle moving freely between two end
walls, for the case of a {\it uniform} initial wavefunction \cite{berry}.
Here his derivation is generalised to 
predict the quantum carpet structure for a broad class of 
initial wavefunctions, and  
in the Appendix it is shown how this 
approach may be further generalised to predict the curved carpet
structures corresponding to {\it arbitrary} one-dimensional potentials.

Schr\"{o}dinger's equation for a free particle of mass $M$ confined 
to the interval $(0,L)$, under the usual boundary conditions that the 
wavefunction vanishes at the endpoints, has the general solution
\begin{eqnarray} \label{sol}
\psi (x,t) & = & (2/L)^{1/2} \sum_{n=1}^{\infty} \psi_n e^{-i\pi n^2 
Vt/L} \sin n\pi x/L  \\ \label{diff}
& = & -i/\sqrt{2L} \left[ \sum_{n=1}^{\infty} \psi_n 
e^{i\phi_{+}(x,t,n)} -  \sum_{n=1}^{\infty} \psi_n 
e^{i\phi_{-}(x,t,n)} \right]  ,
\end{eqnarray}
where $\psi_n$ denotes the $n$-th energy amplitude, $V$ is the 
speed defined in (\ref{vee}), and the phases $\phi_{\pm}$ are given 
by
\begin{equation} \label{phi}
\phi_{\pm}(x,t,n) =  (\pm n x -n^2 Vt)\pi /L  .
\end{equation}

The position amplitude $\psi (x,t)$, and hence the position 
probability distribution $P(x,t) = | \psi (x,t)|^2$, will be 
small if partial cancellation can be arranged between the two 
summations in (\ref{diff}).  This is possible in particular if the term 
corresponding to
summation index $n$ in the first summation partially cancels with the
term corresponding to summation index $n+k$ in the second 
summation, where $k$ is some fixed integer.  Indeed, it is not difficult to 
show that (\ref{diff}) can be rearranged (to within an overall phase 
factor) as 
\begin{eqnarray}
\psi (x,t) & = & (2L)^{-1/2} \sum_{n=1}^{\infty} \left[ \psi_{n+\mid k\mid 
}e^{i\phi_{\pm}(x,t,n+\mid k\mid )} - \psi_n e^{i\phi_{\mp}
(x,t,n)} \right]\nonumber \\ 
\label{gen}
&  + & (2L)^{-1/2} \sum_{n=1}^{\mid k\mid} \psi_n 
e^{i\phi_{\pm} (x,t,n) }  ,
\end{eqnarray}
where the upper (lower) phase subscript is chosen when $k$ is 
positive (negative).
Thus significant cancellation between energy amplitudes can take 
place in the first summation, leading to a small value of $P(x,t)$,
providing that the phase-matching condition
\begin{equation} \label{phase}
\phi_{+}(x,t,n+k) = \phi_{-}(x,t,n) \hspace{1cm} \pmod{2\pi}
\end{equation}
is met for all $n$. 

It is the phase condition (\ref{phase}) which leads to (\ref{berry}).
In particular, note that channel structures in the probability
landscape correspond to  spacetime trajectories
$x(t)$ along which the position probability density is relatively 
low, and hence in particular to trajectories for which the phase
condition (\ref{phase}) is satisfied at all points.  Substituting 
(\ref{phi}) into (\ref{phase}), differentiating 
with respect to time, and comparing consecutive values of $n$, 
yields the condition $\dot{x}(t)=kV$ for the 
slope of such a trajectory.
Substituting $t=0$ in (\ref{phase}) further yields, on comparison
for consecutive values of $n$, condition (\ref{berry}).

Note that while condition (\ref{berry}) thus follows from destructive 
interference between pairs of energy amplitudes in (\ref{gen}),  two further 
conditions are also necessary in general for channel structures to be 
observed.  In particular, substitution of 
(\ref{berry}) in (\ref{gen}) gives the expression
\begin{eqnarray} \label{chan}
\psi (kVt + lL,t) & = & (2L)^{-1/2} \sum_{n=1}^{\infty} (-1)^{nl}
(\psi_{n+\mid k\mid} - \psi_n) e^{-in(n+\mid k\mid )\pi Vt/L} \nonumber 
\\ \label{int}
& + & (2L)^{-1/2} \sum_{n=1}^{\mid k\mid} (-1)^{nl} \psi_n 
e^{-in(n-\mid k\mid )\pi Vt/L}
\end{eqnarray}
for the position amplitude along the trajectory.   
Thus for the trajectory to correspond to  a channel one 
requires further that (i) {\it the differences} $|\psi_n - 
\psi_{n+\mid k\mid}|$ {\it are small} , and (ii) $|k|$ 
{\it itself is sufficiently small}.  These requirements guarantee that 
the first and second summations in (\ref{int}) yield, 
respectively, relatively small contributions to the total amplitude, and
hence to $P(x,t)$.  They hold for 
a wide group of initial wavefunctions, as discussed in section 2.3, 
and hence condition (\ref{berry}) has a wide predictive  power.

\subsection{Channel depths and ridge heights}

Equation (\ref{int}) leads to a simple estimate of channel depth.  In 
particular, the quadratic $n$-dependence of the phases in 
(\ref{int}) implies that they are quasi-random in time, and hence 
the average probabilty density along a given channel can be 
estimated as
\begin{eqnarray} 
\overline{P(kVt + lL,t)} & \approx & (2L)^{-1} \sum_{n=1}^{\infty}
\mid\psi_n - \psi_{n+\mid k\mid}\mid^2 + (2L)^{-1} 
\sum_{n=1}^{\mid k\mid} \mid\psi_n \mid^2 \nonumber \\
\label{pav}
& = & L^{-1} ( 1 - \sum_{n=1}^{\infty} {\rm Re}\{\psi_{n}^{*} 
\psi_{n+\mid k\mid} \} ).
\end{eqnarray}
Note that for slowly varying 
energy amplitudes this expression will typically increase as 
$|k|$ increases, i.e., deep channels correspond to small values of 
$|k|$.  It will be seen in section 3 that this expression is
close to the
exact average depth of the channel.  

A particularly simple example is when the
initial wavefunction is an equally weighted superposition of $N$
consecutive
energy eigenstates, i.e., $\psi_n = 1/\sqrt{N}$ for $M < n \leq M+N$
and $\psi_n = 0$ otherwise, for some $M\geq 0$.  From (\ref{int}) it
follows
that relatively good destructive interference takes place for $|
k|\ll N$, and from (\ref{pav}) that the average density along a
channel is approximately
given by $|k| /(NL)$.  Thus deep channels correspond to small
values of $|k|$, as expected.
Note moreover from (\ref{chan}) that destructive
interference is impossible for $|k|\geq N$, and hence only a
finite number of channels can be observed in this case.  A previous
analytic study of this example for $M=0$
confirms these results \cite{gross}.

{\it Constructive} interference between energy
amplitudes in the first summation in (\ref{gen}) corresponds to adding
$\pi$ to one side of the phase-matching condition (\ref{phase}). It 
follows that constructive interference takes place along trajectories as
per (\ref{berry}), except that the product $kl$ is now restricted to be
{\it odd}.  Thus ``ridges" are predicted along such lines in the
probability landscape, in agreement with numerical observations
\cite{kinzel} - \cite{kap}, \cite{irena}.  
Moreover, the average height of these ridges may
be estimated similarly to channel depths above, yielding an expression
similar to (\ref{pav}) but with subtraction replaced by addition.  Thus
for the above example the average ridge height is predicted to be
$(2N-|k|)/(NL)$ for $|k|\ll N$.

Finally, note that fluctuations of channel depth along a given channel are
constrained to be small, simply because they are bounded by the
relatively small average probability density (which must remain
positive along the channel). In contrast, fluctuations of ridge height
are only bounded by the relatively large average ridge height, 
and indeed are expected to be of the same order from the  
quasi-random nature of the phases appearing in the analogue of (\ref{int})
for constructive interference (where the minus sign in the
first summation is replaced by a plus sign). It is these relatively large
fluctuations which lead to the observed generic ``chopped" nature of 
ridge structures \cite{gross}. 

\subsection{Examples}

The usefulness and limitations 
of the generalised destructive interference approach 
is investigated here via two generic examples.  It will be shown
in particular that channels are expected as per (\ref{berry}) for 
well-localised initial wavepackets; and conversely that destructive
interference fails to account 
for all observed channels in the case of wavepackets with periodically 
spaced energy amplitudes.  

{\it \bf Example (i) Momentum amplitudes and localised wavepackets:}  
The probability landscapes of 
wavefunctions initially well-localised in a one-dimensional box have 
been studied in a number of special cases \cite{gross,stift,kap,irena}. 
Here it is 
shown that the destructive interference approach provides a general 
explanation of channel locations for {\it all} initial wavepackets 
which have  a slowly varying momentum amplitude distribution, and 
for well-localised wavepackets in particular.  

Let $\tilde{\psi}(p)$ denote the momentum amplitude distribution of 
the initial wavepacket, i.e., 
\begin{equation} \label{mom}
\tilde{\psi}(p) = (2\pi\hbar)^{-1/2} \int_{-\infty}^{\infty} 
e^{-ipx/\hbar}\psi(x,0) dx  . 
\end{equation}
It follows from (\ref{sol}) and (\ref{mom}), recalling that $\psi (x,0)$
vanishes outside the interval $(0,L)$, that the energy amplitudes 
$\psi_n$ can be expressed as
\begin{eqnarray} \label{psin}
\psi_n & = & (2/L)^{1/2} \int_{0}^{L} \sin(n\pi x/L) \psi(x,0) dx 
\\ \label{four}
& = & i (\pi\hbar/L)^{1/2} [ \tilde{\psi}(n\pi\hbar/L) - \tilde{\psi}(-
n\pi\hbar/L) ]  .
\end{eqnarray}
From (\ref{chan}) and (\ref{four}) it follows the destructive 
interference approach predicts channels of slope $kV$, as per 
(\ref{berry}), {\it providing that the momentum amplitude 
distribution } $\tilde{\psi}(p)$ {\it varies sufficiently slowly over any 
range of length} $|k|\pi\hbar/L$.  

In particular, if the initial wavepacket $\psi(x,0)$ is smooth and 
well-localised within the box then the momentum amplitude distribution 
will typically be broad, with root mean square variance $\Delta p$ say.  
Thus channels of slope $kV$ are predicted for 
\begin{equation} \label{loc}
 |k|\pi\hbar/L \ll \Delta p . 
\end{equation}
Noting that the Heisenberg uncertainty relation implies 
that $\Delta p \geq \hbar/(2\Delta x)$, one in particular expects to see 
channels of slope $kV$ for all smooth localised wavepackets such that
$|k| \ll L/(2\pi\Delta x)$.
These general predictions are well borne out by the approximately Gaussian 
initial wavepackets studied previously \cite{kinzel}, 
\cite{stift} - \cite{irena}.

{\it \bf  Example (ii) Periodically spaced energy amplitudes:}  
The second example to 
be considered here is the case where the non-zero energy amplitudes 
of the wavefunction are periodically spaced, i.e.,
\begin{equation} \label{per}
\psi_n = 0 {\rm \: for \:} n \neq r \bmod p
\end{equation}
for some period $p$ and fixed integer $r$.

The case $p=1$ is trivial, with channels predicted as per (\ref{berry}) 
just as before.  The case $p=2$ corresponds to initial 
wavefunctions which are either symmetric or antisymmetric about 
$x=L/2$, as $r$ is odd or even respectively, as can be directly seen from the
form of the wavefunction in (\ref{sol}).  
The case $p=3$ will be seen to be of special
significance, providing an example where the
destructive interference approach breaks down.

From (\ref{per}) it is seen that amplitudes  $\psi_n$ and $\psi_{n+\mid
k\mid}$ can only destructively interfere in the first summation in
(\ref{gen}) when $k$ is a multiple of the period $p$ and $n = r \bmod
p$.  Accordingly, substituting
$pk$ for $k$ and $pn+r$ for $n$ in the phase-matching condition
(\ref{phase}) leads to the modification 
\begin{equation} \label{pberry}
x = kpVt + lL/p , \hspace{1cm} (k + 2r/p)l = {\rm even \; integer}
\end{equation}
of (\ref{berry}), for trajectories which correspond to channels in the
probability landscape.
Condition (\ref{pberry}) is equivalent to (\ref{berry}) for $p=1$.
More generally the predicted channel locations depend upon both 
$p$ and $r$, have slopes which are multiples of $pV$, 
and intersect the $x$-axis at multiples of $L/p$.  

For example, for $p=2$ and $r=1$ (symmetric initial wavefunctions), 
it follows that channels correspond to the trajectories $x=2kVt+lL/2$,
such that $(k+1)l$ is even.  This prediction is equivalent to
Eq. (52) of \cite{berry}, obtained there
for the special case of an initially uniform wavefunction.

The case $p=3$ provides an example which demonstrates an incompleteness of
the destructive interference approach.  In particular, 
consider an initial wavefunction given by the 
equally-weighted superposition 
\begin{equation} \label{fig1}
\psi (x,0) = N^{-1/2} \sum_{n=0}^{N-1} \sin [(3n+1)\pi x/L ] .
\end{equation}
This corresponds to $p=3$ and $r=1$ in (\ref{per}), and hence 
from (\ref{pberry}) {\it no} channels starting
from $x=L/3$ are predicted.  
However, the density plot of the corresponding probability landscape for
this example, shown in Figure 1 for $N=20$, shows that
a channel of slope $-3V$ is in fact 
associated with this starting point.  The appearance of this unexpected
channel will be explained in the following section.

\section{Travelling wave approach} 

\subsection{Structure function decomposition}

It has been seen that the destructive interference approach successfully
predicts a number of generic properties of quantum carpets, including
the locations of channel and ridge structures and their corresponding
average depths and heights. It does not, however, yield information on
the shapes and widths of these structures; does not predict {\it all} observed
structures (Figure 1); and is in any case 
limited in applicability to 
wavefunctions with energy amplitudes which are slowly varying
over ranges of length $|k|$.  Thus a new, more general approach is
desirable.

To introduce such an approach, it is first convenient to rewrite the
wavefunction in (\ref{sol}) in the form \cite{irena}
\begin{equation} \label{sol2}
\psi (x,t) = -i(2L)^{-1/2} \sum_{n=-\infty}^{\infty} \psi_n
e^{i(nx - n^2 Vt)\pi /L}  ,
\end{equation}
where one extends the energy amplitude coefficients $\psi_n$ to negative
values of $n$ via the definition
\begin{equation} \label{extend}
\psi_{-n} = - \psi_n  .
\end{equation} 

From (\ref{sol2}) one immediately has
\begin{equation} \label{psim}
P(x,t) = (2L)^{-1} \sum_{m,n=-\infty}^\infty \psi_{m}^{*} \psi_n
e^{-i(m-n)[x-(m+n)Vt]\pi /L}
\end{equation}
for the position probability distribution.  It may be seen that each
term in this summation either has no spacetime dependence ($m=n$), or is
a plane wave with velocity $(m+n)V$ ($m\neq n$).  Collecting the $m=n$ terms
into a constant background term, and grouping terms corresponding
to waves propagating at velocity $kV$, one obtains
the decomposition
\begin{equation} \label{trav}
P(x,t) = L^{-1} \left[ 1 + \sum_{k=-\infty}^\infty S_k
\left(\frac{x-kVt}{L}\right)\right]
\end{equation}
of the probability landscape, where the ``structure 
functions" $S_k$ are given by
\begin{equation} \label{struc}
S_k (z) = \frac{1}{2} \sum_{m=-\infty}^{\infty} \psi_{m}^{*} \psi_{k-m}
e^{-i(2m-k)\pi z} - \frac{1}{2} |\psi_{k/2}|^2 . 
\end{equation}
Here $\psi_{k/2}$ is defined to be zero when $k$ is odd.

Equation (\ref{trav}) is the travelling wave decomposition referred to
in the Introduction.  The first term, $L^{-1}$, is a constant background
term for the probability distribution, and the structure function $S_k$ is
a real travelling wave travelling at velocity $kV$. Note that $S_k$ is
defined on the entire real axis, and satisfies the relations
\begin{equation} \label{repeat}
S_k (z+1) = (-1)^k S_k (z)  ,\;\;\;\;\;\; S_{-k}(z) = S_{k} (-z) ,
\end{equation}
where the latter of these follows using (\ref{extend}).

The travelling wave decomposition (\ref{trav}) provides a physical
picture for the generation of the probability landscape $P(x,t)$, as the
superposition of waves of discrete velocities propagating in spacetime
against a constant background probability.  However, the primary
usefulness of this decomposition arises from its relationship to the
statistical properties of the probability landscape.  

In particular, from (\ref{struc}), the time average of $S_k (z+\alpha t)$ is
equal to $S_k (z)$ if $\alpha =0$ and vanishes otherwise.
It follows via (\ref{trav}) that the average of $P(x,t)$ along the linear
trajectory $x(t) = x_0 + kVt$ is given by
\begin{equation} \label{aver}
\overline{P(x_0 + kVt, t)} =  L^{-1} [1 + S_k (x_0 /L) ]  
\end{equation}
for any integer k, and by $L^{-1}$ when $k$ is not an integer.  Thus the
average distribution along any direction in spacetime involves at most
{\it one} structure function.

\subsection{Locations, shapes and widths of linear structures}

From (\ref{aver}) it is seen that $S_k (z)$ contains all information
pertaining to the average properties of linear structures of slope $kV$
in a quantum carpet.  In particular, channel/ridge structures
are associated with those trajectories $x=x_0+kVt$ for which $x_0/L$ 
corresponds to minima/maxima respectively of $S_k$.
Moreover, the average shape of these stuctures corresponds to the shape
of the associated structure function.

For example, $1+S_{-3}(z)$ is plotted in Figure 2
for the wavefunction (\ref{fig1}) (with $N=20$).  The deep maximum in Figure
2 in the vicinity of $z=1/3$ implies, via relation (\ref{aver}), that a
well-defined channel structure of slope $-3V$ crosses the $x$-axis at
$L/3$.  This channel is precisely  the unexpected channel observed in
Figure 1, which was not predicted by the destructive interference
approach of section 2.  Note also from Figure 2 that a ridge structure
immediately parallel to the left of this channel is also predicted, as
indeed may also be observed in Figure 1.  Since the structure
functions in (\ref{struc}) can be trivially evaluated as sums of
geometric series for this example, the average properties of all 
linear structures can in fact be calculated analytically via
(\ref{aver}) if desired. 

In Figure 3 the travelling wave decomposition (\ref{trav}) is
directly illustrated, again for the wavefunction (\ref{fig1}) with
$N=20$, where only terms with $|k|\leq5$ have been included in the
summation.  Comparison with Figure 1 demonstrates that all linear
carpet structures with slope less than or equal to $5V$ in magnitude are
reproduced in Figure 3.  

Note finally that one can make use of (\ref{aver}) to define the average
horizontal width of a given channel or ridge
structure.  In particular, suppose such a structure corresponds to a 
trajectory passing through the $x$-axis at $x_0$. Define $x_+$ and
$x_-$ to be the points to the right and left respectively of $x_0$ for which
the average probability distribution first becomes equal to the
background probability $L^{-1}$.  The difference $x_+ - x_-$ is then a
natural measure of the horizontal width of the structure.  
From (\ref{aver}) these
points correspond to the zeroes of the structure function lying either
side of $x_0 /L$, where $x_0/L$ itself corresponds to a minimum or maximum
of the structure function.  Note for a structure of slope $kV$ that a 
corresponding measure of width in the direction {\it perpendicular} 
to the structure is 
obtained by dividing the horizontal width by $(1+k^2V^2)^{1/2}$.

\subsection{Connections between the approaches}

Since from (\ref{aver}) the structure functions are real, one can
rewrite (\ref{struc}) as
\begin{eqnarray}
4 S_k (z) + 2|\psi_{k/2}|^2 & = & 2 \sum_{m=-\infty}^{\infty} {\rm Re} \left\{ 
\psi_{m}^{*} \psi_{k-m} e^{-i(2m-k)\pi z} \right\} \nonumber \\
& = & \sum_{m=-\infty}^{\infty} \left\{ |\psi_{k-m} + \psi_m
e^{i(2m-k)\pi z}|^2 - |\psi_{k-m}|^2 - |\psi_m |^2 \right\} .\nonumber
\end{eqnarray}
Replacing $m$ by $m+k$ in the first term of this sum and using
(\ref{extend}) then gives an alternate formula for the average
probability distribution in (\ref{aver}): 
\begin{equation} \label{comp}
\overline{P(x_0 + kVt, t)} = (4L)^{-1} \left[ \sum_{m=-\infty}^{\infty} 
\left|\psi_m - \psi_{m+k}e^{i(2m+k)\pi x_0 /L}\right|^2 - 
2|\psi_{k/2}|^2 \right] .
\end{equation}
 
The above expression clearly indicates a link between the average
probability distribution and
interference of the amplitudes $\psi_m$ and
$\psi_{m+k}$.  Further, this interference is seen to be {\it destructive}
when $x_0=lL$ for some integer $l$ with $kl$ even, thus recovering
condition (\ref{berry}) for channel structures.  
The destructive interference approach is thus a special case of the
travelling wave approach, where condition (\ref{berry}) corresponds to
particular minima of the structure functions (at $z=l$) for the case of
slowly varying energy amplitudes.

Finally, (\ref{extend}) can be used to rewrite (\ref{comp}) in terms of energy
amplitudes $\psi_m$ with $m>0$:
\begin{equation} \label{comp2}
\overline{P(kVt + lL, t)} = L^{-1} {\rm Re}\left\{ 1 - \sum_{m=1}^{\infty}\psi_{m}^{*} 
\psi_{m+|k|} + \frac{1}{2}\sum_{m=1 \; (m\neq |k|/2)}^{|k|-1} \psi_{m}^{*}  
\psi_{|k|-m} \right\} 
\end{equation}
when $kl$ is an even integer. Thus the approximation in (\ref{pav}) is
seen to be quite good for small $|k|$, and indeed is exact for $|k|$
equal to 1 or 2. 

\section{Properties of structure functions}

\subsection{Wigner function form}

Expression (\ref{struc}) for the structure function $S_k$ is not
always convenient to use.  It generally involves an infinite summation,
and the contributing energy amplitudes are not always easily
calculated.  It would therefore be useful to have a formula for $S_k$
which is directly related to the initial wavefunction $\psi (x,0)$.

To obtain such a formula, first define the normalised wavefunction
$\phi (x)$ on the interval $(-L,L)$ by
\begin{eqnarray} \label{phix}
\phi (x) & = & 2^{-1/2} [ \psi(x,0) - \psi(-x,0) ]\\
\label{phi2} & = & L^{-1/2}/(2i) \sum_{n=-\infty}^{\infty} \psi_n e^{in\pi
x/L}   ,
\end{eqnarray}
where the second line follows via (\ref{sol2}) and (\ref{extend}).  Thus
$\phi(x)$ is an antisymmetric extension of the initial wavefunction
$\psi(x,0)$.  Now let $W_\phi (x,p)$ denote the Wigner function of
$\phi(x)$ \cite{wig}:
\begin{equation} \label{wig}
W_\phi (x,p) = (\pi\hbar)^{-1} \int_{-\infty}^{\infty} dy \phi^* (x-y)
\phi(x+y) e^{-2ipy/\hbar}  .
\end{equation}
For $0\leq x<L$ one then has, as shown further below, the remarkable relation
\begin{equation} \label{remark}
S_k(x/L) = \pi\hbar \left[ W_\phi (x,p_k) + 
(-1)^k W_\phi (x-L,p_k)\right]
- |\psi_{k/2}|^2/2  ,
\end{equation}
connecting $S_k$ with $W_\phi$, where
\begin{equation} \label{pk}
p_k = \pi\hbar k/(2L)   = kMV.
\end{equation}
This expression can be trivially extended to evaluate $S_k(z)$ for all
values of $z$ via the first of the relations in (\ref{repeat}).

As will be seen below, (\ref{remark}) provides a
very convenient method for evaluating structure functions, and also for
analysing the dependence of quantum carpets on various
properties of the initial
wavefunction.  Expansions of the probability
distribution $P(x,t)$ in terms of Wigner functions have
been previously obtained \cite{stift,marz}.  However, these expansions
lead only to expressions for $S_k$ involving {\it infinite} sums of
Wigner functions, in contrast to (\ref{remark}). 
The latter may be derived by substituting (\ref{phix}) and (\ref{pk}) into
(\ref{wig}) and recalling that $\phi(x)$ vanishes for $|x|\geq L$ by
definition, thus yielding
\begin{eqnarray*}
4\pi\hbar LW_\phi (x,p_k) & = & \sum_{m,n} \psi_{m}^{*}\psi_n
e^{-i(m-n)\pi x/L}\int_{-M(x)}^{M(x)} dy e^{i(m+n-k)\pi y/L} \\
& = & 2\sum_{m+n\neq k} \psi_{m}^{*}\psi_n e^{-i(m-n)\pi x/L}
\frac{\sin[(m+n-k)\pi M(x)/L]}{(m+n-k)\pi /L} \\
&  & \mbox{} + 2 \sum_m \psi_{m}^{*}\psi_{k-m}e^{-i(2m-k)\pi x/L} M(x)  , 
\end{eqnarray*}
where
\begin{equation} \label{mx}
M(x) := \max \{ L-|x|, 0\}  .
\end{equation}
Substitution into the righthand side of (\ref{remark}) gives
(\ref{struc}) as required.

\subsection{Localisation and symmetry effects}

To see how localisation and symmetry properties of the initial
wavefunction directly affect the structure functions, one may substitute
(\ref{phix}) and (\ref{wig}) into (\ref{remark}), to obtain
\begin{eqnarray} 
S_k (x/L) & = & (\pi\hbar /2) \left[ W_\psi (x,p_k) + (-1)^k W_\psi
(L-x,-p_k)\right] \nonumber\\ \label{wigpsi}
&  & \mbox{} - (1/2)\left[ I_\psi (x,p_k) + |\psi_{k/2}|^2 \right]
\end{eqnarray}
for $0\leq x<L$, where $W_\psi$ denotes the Wigner function of $\psi
(x,0)$, and the ``interference" term $I_\psi (x,p_k)$ is given by
\begin{equation} \label{int2}
I_\psi (x,p_k) = \left\{ \begin{array}{ll}
\int_{-\infty}^{\infty}dy\psi^*(y+x,0)\psi (y-x,0)e^{i\pi ky/L} 
+ c.c., & 0\leq x\leq L/2\\
(-1)^k I(L-x, -p_k) , & L/2 < x < L .\end{array} \right.
\end{equation}

Now, for example, suppose that the initial wavefunction $\psi(x,0)$ is
well-localised about some point $x_0\neq 1/2$. The corresponding Wigner
function $W_\psi (x,p)$ will then be
similarly localised, and hence from (\ref{aver}) and
(\ref{wigpsi}) one predicts (i)
a ridge structure associated with the trajectory $x_0 + kVt$; and (ii) a
ridge/channel structure associated with the trajectory $L-x_0 +kVt$
for even/odd values of $k$. An example verifying this prediction is given in
section 5.2 below.

It is also of interest to consider the cases of trajectories of the
form $x=lL + kVt$ (arising from the destructive interference approach in
section 2).  From (\ref{repeat}),
(\ref{wigpsi}) and (\ref{int2}) one has
\begin{eqnarray} 
S_k(l) & = & (-1)^{kl} S_k(0) \nonumber\\ \label{corn} & = & 
(-1)^{kl+1}\left[ \int_0^L dx P(x,0) \cos(\pi kx/L) 
+ |\psi_{k/2}|^2/2 \right] ,
\end{eqnarray}
where $P(x,0)$ is the initial position probability distribution.  The
average probability density along these trajectories thus only depends on the 
phase structure of $\psi(x,0)$ via $\psi_{k/2}$
(which is zero for odd values of $k$ and typically small in general).

If initial distribution $P(x,0)$ is well-localised
within a region $(x_0-\delta x, x_0+\delta x)$ such that
\begin{equation} \label{delta}
\delta x/L \ll |k|^{-1} , 
\end{equation}
then from (\ref{corn}) the average probability density
along $x=lL+kVt$ is essentially determined by the properties of the cosine
function $\cos \pi kx_0/L$.
For example, if $x_0$ satisfies 
\begin{equation} \label{max}
kx_0/L = j   
\end{equation}
for some integer $j$, then 
from (\ref{aver}) and  (\ref{corn}) one has 
\begin{equation} \label{local}
\overline{P(lL + kVt, t)} \approx (1 - (-1)^{j+kl} )/L,  
\end{equation}
corresponding to channel and ridge structures as $j+kl$ is odd and even
respectively  (the $\psi_{k/2}$ term has been ignored in 
(\ref{local}), as from (\ref{psin}) it is typically negligible for
localised wavefunctions).  
Similarly, if $j$ is replaced by $j+1/2$ in (\ref{max}) then the
$(-1)^{kl+j}$ term in (\ref{local}) vanishes, and any linear 
structure associated with the trajectory is suppressed.

Finally, rather than supposing the initial distribution to be 
well-localised, consider instead the case where $P(x,0)$  has an approximate
reflection symmetry about some point $x^*$, i.e.,
\begin{equation} \label{symm}
P(x^*+x,0) \approx P(x^*-x,0)  .
\end{equation}
Noting that $P(x,0)$ vanishes outside $(0,L)$ one then has from (\ref{corn})
that 
\begin{eqnarray}
(-1)^{kl}S_k (l) & \approx & -\int_0^\infty dx P(x^*+x,0)[
\cos\pi k(x^*+x)/L
\nonumber\\ & & \mbox{} + \cos\pi k(x^*-x)/L ] -
|\psi_{k/2}|^2/2
\nonumber\\ \label{supp}
& = & -2\cos\frac{\pi kx^*}{L} \int_0^{\infty}\!\! dx P(x^*+x,0) 
\cos\frac{\pi kx}{L} -\frac{1}{2}|\psi_{k/2}|^2 .
\end{eqnarray}
Thus the structure function is modulated by the cosine function
$\cos\pi kx^*/L$.
Linear structures associated with the trajectories $x=kVt + lL$
are therefore enhanced when $kx^*/L=j$ for some integer $j$, and suppressed when
$kx^*/L=j+1/2$.  Examples  of such
enhancement/suppression are given in the following section.  

\section{Examples}

\subsection{Uniform initial wavefunction}

The case $\psi(x,0)=L^{-1/2}$ on (0,L) was considered by Berry
\cite{berry}, who showed that the corresponding probability landscape
was a fractal and explained the observed channel structures via
destructive interference (Example (ii) of section 2.3 above).  For
this case it follows from (\ref{psin}) and (\ref{extend}) that
\begin{equation} \label{coeff}
\psi_n = 2\sqrt{2}/(n\pi)
\end{equation}
for odd $n$, with $\psi_n$ vanishing for even 
$n$.  The structure function $S_{2k}(z)$ can be evaluated via any of
(\ref{struc}), (\ref{remark}) or (\ref{wigpsi}), to give 
\begin{equation} \label{unif}
S_{2k}(z) = 2(\pi k)^{-1} \sin [2\pi k\min\{z,1-z\} ] -\delta_{k0} -
|\psi_k|^2/2 
\end{equation}
for $0\leq z<1$, where $\delta_{kl}$ is the Kronecker delta and 
$\sin[x]/x$ is evaluated as 1 for $x=0$.  
As usual this may be extended to other values of
$z$ via (\ref{repeat}).

In Figure 4 (\ref{unif}) is plotted for $k=1$, 2 and 3.  It is seen that
while channels are associated with trajectories $x=2kVt+lL/2$ for even 
values of $(k+1)l$, as predicted in Example (ii) of section 2.3, these
channels are relatively broad and shallow.  Indeed the average probability
density along these channels follow from (\ref{repeat}), (\ref{aver}) and
(\ref{unif}) as just the background probability $1/L$ for even 
$k$, and $[1-4/(\pi k)^2]/L$ for odd $k$.  Channel visibility thus
decreases rapidly as $|k|$ increases.

Since the initial probability distribution is symmetric 
about $x^*=L/2$ for this example,
one expects from (\ref{supp}) that channels with slopes of $(2k+1)V$
are suppressed.  Indeed, as is most easily seen from (\ref{comp}) 
(recalling $\psi_n$ vanishes for even values of $n$), one
finds $S_{2k+1}(z)\equiv 0$. Hence {\it no} linear structures of slope
$(2k+1)V$ are predicted, and (\ref{trav}) and (\ref{unif}) then
yield the surprisingly simple decomposition
\begin{equation} \label{beaut}
P(x,t) = 2\sum_{k=-\infty}^{\infty} (\pi kL)^{-1} \sigma(x-kVt)
\sin [2\pi k(x-kVt)/L] + [1-2\sigma(x)]/L
\end{equation}
of the fractal probability landscape, where $\sigma(x)$ is defined to be
$+1$, 0, and $-1$ as the fractional part of $x/L$ is respectively 
less than, equal to, or
greater than $1/2$.

\subsection{Gaussian initial wavefunctions}

Consider now the inital wavefunction
\begin{equation} \label{gauss}
\psi(x,0) = K (2\pi\sigma^2)^{-1/4} e^{-(x-\overline{x})^2/
(4\sigma^2)} \, e^{i\overline{p} x},
\end{equation}
where $0<x<L$ and $K$ is a normalisation constant.  It will be assumed
that $\sigma\ll \overline{x}$, $L -\overline{x}$, i.e., 
that the wavefunction is
well-localised on the interval.  One may then make the extremely good
approximation $K=1$, so that 
$\psi(x,0)$ is effectively a Gaussian
wavepacket centred at $\overline{x}$ with average momentum $\overline{p}$.

It is easiest to evaluate the structure functions for this case via
(\ref{wigpsi}) and (\ref{int2}), where one makes the (again extremely good)
approximation that (\ref{gauss}) can be extended over the entire
$x$-axis.  Performing the resulting Gaussian integrals then yields
\begin{eqnarray} 
S_k\left( \frac{x}{L}\right)
 & \approx & \frac{1}{2}\left[e^{\frac{-(x-\overline{x})^2}
{2\sigma^2}}
e^{\frac{-2\sigma^2(p_k-\overline{p})^2}{\hbar^2}} + (-1)^k
[e^{\frac{-(L-x-\overline{x})^2}{2\sigma^2}}
e^{\frac{-2\sigma^2(p_k+\overline{p})^2}{\hbar^2}} \right]\nonumber\\
\label{gstruc} & & \mbox{} -
\left\{ \begin{array}{ll}
e^{\frac{-x^2}{2\sigma^2}}e^{\frac{-\pi^2 k^2 \sigma^2}{2L^2}}
\cos \frac{2}{\hbar}(p_k\overline{x}-\overline{p}x) , 
& 0\leq x\leq \frac{L}{2}\\
(-1)^k e^{\frac{-(L-x)^2}{2\sigma^2}}e^{\frac{-\pi^2 k^2 \sigma^2}{2L^2}}\cos
\frac{2}{\hbar}
(p_k\overline{x}+\overline{p}(L-x)) , & \frac{L}{2} < x < L .\end{array} \right.
\end{eqnarray}
The term $-(1/2)|\psi_{k/2}|^2$ in (\ref{wigpsi}) has been ignored, as it 
vanishes for odd $k$ and from (\ref{mom}) and (\ref{four}) 
is only of order $\sigma/L\ll 1$ for even $k$.

The structure function $S_1(z)$ is plotted in Figure 5 with 
$\overline{x}=L/3$ and $\sigma=L/40$, 
for the cases $\overline{p}=0$ (dashed line) and
$\overline{p}=15\pi\hbar/L$ (solid line). In the latter case it is seen that
there is a deep channel associated with the trajectory $x=Vt$, flanked
by a high ridge on the left and a moderate ridge on the right.  Further,
from (\ref{repeat}) one has $S_{-1}(1+z)=S_1(1-z)$, and hence there
is, conversely, a high ridge associated with the trajectory $x=L-Vt$,
flanked by a deep channel on the right and a moderate channel on the
left.  These predicted features following from Figure 5 may be directly
observed in Figure 6, where the probability landscape corresponding to
initial wavefunction (\ref{gauss}) is plotted for $\overline{x}=L/3$ and
$\overline{p}=15\pi\hbar/L$.  The structure functions thus accurately predict
the shapes of the observed linear structures.

For the $\overline{p}=0$ case (dashed line) in Figure 5 it is seen that
as well as channel and ridge structures at the endpoints, there is a
ridge associated with the trajectory $x=L/3+Vt$, and a channel
associated with the trajectory $x=2L/3+Vt$.  These latter structures
correspond to the Wigner functions $W_\psi(x_0,p_k)$ and
$W_\psi(L-x_0,p_k)$ discussed in section 4.2 for localised
wavefunctions, and can again be observed in a plot of the probability
landscape.  The corresponding structures for the
$\overline{p}=15\pi\hbar/L$ case are observed to be much less pronounced.
This is because $W_\psi(\overline{x},p)$ is localised about
$p=\overline{p}$, and is thus relatively small at $p=p_1$. 

Gaussian initial wavefunctions also provide an example of the prediction
in (\ref{local}) for well localised initial wavefunctions.  In
particular, from (\ref{gstruc}) (recalling the assumption $\sigma\ll 
\overline{x}$, $L-\overline{x}$), one finds that 
\begin{equation} \label{cosi}
S_k(0)\approx -e^{\frac{-\pi^2 k^2 \sigma^2}{2L^2}}\cos\pi k\overline{x}/L.
\end{equation}
Hence (\ref{local}) follows whenever conditions (\ref{delta}) and
(\ref{max}) are satisfied 
(with $\delta x = \sigma$ and $x_0=\overline{x}$). Further, channels and
ridges associated with trajectories $x=kVt+lL$ are seen to be suppressed if
$\pi k\overline{x}/L=j+1/2$ for some integer $j$. Since 
Gaussian initial wavefunctions are not only localised about
$\overline{x}$, but are approximately symmetric about $\overline{x}$,
one may also obtain (\ref{cosi}) from (\ref{supp}), with
$x^*=\overline{x}$. 

\section{Periodic gratings and rigid rotators}

Consider now a plane wave of wavelength $\lambda$ incident
on a one-dimensional periodic
grating, of period $2L$ in the $x$ direction.  In the paraxial
approximation the amplitude of the diffracted light then has the general
form 
\begin{equation} \label{diffract}
\hat{\phi}(x,z) = (2L)^{-1/2} \sum_{n=-\infty}^{\infty} \hat{\phi}_n
e^{i(nx-n^2Vz)\pi /L} ,
\end{equation}
where $z$ measures distance propagated perpendicularly to the grating,
the Fourier coefficients $\hat{\phi}_n$ are determined by the initial
diffracted wave $\hat{\phi}(x,0)$, and $V=\lambda /(4L)$.

Clearly (\ref{diffract}) is formally similar to (\ref{sol2}) for a
quantum particle moving freely between two endwalls, and indeed
analogues of quantum carpets in the grating context have been previously
observed
and investigated \cite{klein,boden}.  The significant differences
between (\ref{diffract}) and (\ref{sol2}) are
(i) $\hat{\phi}(x,z)$ is periodic on the entire $x$ axis whereas
$\psi(x,t)$ vanishes outside the interval $(0,L)$; and (ii) the
coefficients $\hat{\phi}_n$ need not be antisymmetric as per (\ref{extend}).

Many of the results of sections 3 and 4 may be directly translated into
the grating context.  For example, the light intensity distribution
$I(x,z)=|\hat{\phi}(x,z)|^2$ has the travelling wave decomposition
\begin{equation} \label{inten}
I(x,z) = (2L)^{-1}\left[ I_0 + \sum_{k=-\infty}^\infty \hat{S}_k
\left(\frac{x-kVz}{L}\right) \right]
\end{equation}
in analogy to (\ref{trav}), where $I_0$ is the integrated light
intensity per period of the grating and $\hat{S}_k$ is given by
(\ref{struc}) with $\psi_n$ replaced by $\sqrt{2}\hat{\phi}_n$.  Similarly,
the analogue of (\ref{aver}) is
\begin{equation} \label{aver2}
\overline{I(x_0+kVz,z)} = (2L)^{-1} [I_0 + \hat{S}_k(x_0/L)],
\end{equation} 
and thus the structure function $\hat{S}_k$ determines the average
locations, shapes, etc of relatively dark and bright
structures of slope $kV$ in the $x$-$z$ plane.
The average intensities in (\ref{aver2}) are precisely the ``lane
contrast functions" first defined in \cite{boden} for predicting the
locations of dark structures.

While one may also write down a result analogous to
(\ref{remark}), an {\it alternative} Wigner function expression for the
structure functions is more useful in the grating context.  In
particular, define the {\it periodic}
Wigner function
\begin{equation} \label{newwig}
W_{\hat{\phi}}(x,p) = \pi^{-1} \int_{-L}^L dy \hat{\phi}^*(x-y,0)
\hat{\phi}(x+y,0)e^{-2ipy}. 
\end{equation}
This differs from (\ref{wig}) in that $\hat{\phi}$ does {\it not}
vanish outside the interval $(-L,L)$.  One may then show, similarly to
the proof of (\ref{remark}), that
\begin{equation} \label{newst}
\hat{S}_k(x/L) = \pi W_{\hat{\phi}}(x,\pi k/(2L)) - |\hat{\phi}_{k/2}|^2 ,
\end{equation}
where $\hat{\phi}_{k/2}$ is defined to vanish for odd $k$.

As an example of (\ref{newst}), suppose that a plane wave is incident on
a sinusoidal phase grating, as studied in \cite{boden}, i.e.,
\begin{equation}
\hat{\phi}(x,0) = (I_0/2L)^{1/2} \, e^{i\alpha\cos (\pi x/L)} .
\end{equation}
Substitution into (\ref{newwig}) and using the standard Bessel integral 
\begin{equation}
\int_0^\pi d\theta \cos [k\theta - a\sin\theta ] =\pi J_k(a)
\end{equation}
then yields, via (\ref{aver2}) and (\ref{newst}),
\begin{equation} \label{examp}
\overline{I(x_0+kVz,z)} = I_0/(2L) \left[ 1+(-1)^k J_k(2\alpha\sin[\pi
x_0/L]) - |\hat{\phi}_{k/2}|^2\right] .
\end{equation}
Here $|\hat{\phi}_{k/2}|$ is zero for odd $k$, and equal to 
$J_{k/2}(\alpha)$ for even $k$ \cite{boden}.

The last term of (\ref{examp}) provides a correction to equation (25) of 
\cite{boden},
where the average intensity along the trajectory $x=x_0+kVz$ was 
calculated for the same example.  More generally, (\ref{newst}) provides a
convenient way to calculate the structure function directly from the initial
amplitude $\hat{\phi}(x,0)$, and (\ref{inten}) provides a simple
decomposition of the intensity distribution as a sum of travelling waves. 

The above results may be immediately translated into the
context of the two-dimensional quantum rigid rotator, with Hamiltonian
$H={J_z}^2/(2I)$.  In particular, $\hat{\phi}(\theta,t)$ may be
interpreted
as the time-dependent phase amplitude of the rotator, 
where one takes $L=\pi$ and $V=\hbar^2/(2I)$.  The carpet structures
for this case lie on the cylinder generated by the phase and time
coordinates $\theta$ and $t$.

\section{Conclusions}

The carpet structure of a quantum particle confined between two endwalls
has been greatly elucidated by the travelling wave decomposition
(\ref{trav}) of the probability density, and the details afforded by
this decomposition are seen to go well beyond the ambit of the
destructive interference approach.  In particular, the
structure functions $S_k$ in (\ref{trav}) provide {\it all} information
pertaining to the average properties of linear structures in the
probability landscape.

The structure functions may be evaluated either from the energy
amplitudes of the wavefunction, via (\ref{struc}); or from the
initial wavefunction, via (\ref{wigpsi}).  The latter formula permits
the dependence of generic features of carpet patterns to be directly
determined from corresponding properties of $\psi(x,0)$.  Similarly, for
the periodic grating and quantum rigid rotator one may use (\ref{newst})
to directly evaluate structure functions from $\hat{\phi}(x,0)$.

One advantage remaining to the destructive interference approach is that
it may be generalised to predict the carpet structure of semiclassical
wavepackets moving in arbitrary one-dimensional potentials (see
Appendix).  It is hoped that a corresponding generalisation of the
travelling wave decomposition (\ref{trav}) can be found. 

{\bf Acknowledgements}

We are grateful to Professor Berry for a copy of \cite{boden} prior to
publication.  MH acknowledges the support of the Alexander von Humboldt
Foundation.
\newpage
\appendix{{\bf Appendix}}

It will be briefly indicated here how the destructive interference
approach of section 2 may be generalised to predict the {\it curved} carpet
structures arising in the probability landscapes of one-dimensional
particles moving in general potentials \cite{stift,kap,loinaz}.  
For semi-classical
wavepackets the analysis is very similar to that of subsection 2.1, and
the results agree with those of Kaplan et al \cite{kap} obtained by
consideration of degeneracies of the propagator for $P(x,t)$.

Now, for high energies the $n$-th energy eigenstate of a particle moving
in a one-dimensional potential $U(x)$ with two classical turning points
is well approximated as \cite{davy}
\begin{equation} \label{wkb}
\psi_n (x) = 2[M/T_n p_n (x)]^{1/2} \sin [\hbar^{-1}\int_{x_n}^{x} p_n(x)
dx + \pi /4 ] ,
\end{equation}
where $M$ is the particle mass; $T_n$, $p_n (x)$ and $x_n$ are the
period, momentum and leftmost turning point respectively of a
corresponding classical orbit of energy $E_n$; and $E_n$ is
implicitly defined by the Bohr-Sommerfeld rule
\begin{equation} \label{bohr}
\oint p_n (x) dx = (n+1/2)2\pi\hbar  .
\end{equation}

The appearance of the sine function in (\ref{wkb}) is analogous to that
in (\ref{sol}), and for an initial superposition $\psi(x,0) =
\sum c_n\psi_n (x)$ of such
states one can easily obtain the analog of (\ref{gen}):
\begin{eqnarray}
\psi (x,t) & = &  \sum_{n=1}^{\infty} \left[ A_{n+\mid k\mid
}(x) e^{i\phi_{\pm}(x,t,n+\mid k\mid )} - A_n (x) e^{i\phi_{\mp}
(x,t,n)} \right]\nonumber \\
\label{arb}
&  + &  \sum_{n=1}^{\mid k\mid} A_n (x)
e^{i\phi_{\pm} (x,t,n) }  ,
\end{eqnarray}
where
\begin{eqnarray} \label{amp}
A_n(x) & = &  [M/T_n p_n (x)]^{1/2} c_n   ,\\ \label{arbphi}
\phi_\pm (x,t,n) & = & \pm \hbar^{-1} \int_{x_n}^x p_n(x) dx
-\hbar^{-1} E_n t  \pm \pi /4  .
\end{eqnarray}
As in (\ref{gen}), the upper (lower) phase subscript is chosen in
(\ref{arb}) when $k$ is positive (negative), and an overall phase factor
has been dropped.

Destructive interference in the first summation in (\ref{arb}), giving
rise
to a channel structure, can take place along a spacetime trajectory $x_k
(t)$ if the phase condition (\ref{phase}) holds for $\phi_\pm
(x,t,n)$ defined in (\ref{arbphi}) (providing that the amplitude
functions
$A_n(x)$ in (\ref{amp}) vary sufficiently slowly with $n$).
Substituting
(\ref{arbphi}) into (\ref{phase}) and differentiating with respect to
$t$
yields the condition
\begin{equation} \label{traj}
\frac{dx_k(t)}{dt} = \pm \frac{E_{n+\mid k\mid} -E_n}{p_{n+\mid
k\mid}(x_k) +
p_n(x_k)}
\end{equation}
for the trajectory $x_k(t)$, where the $+/-$ sign is chosen according to
whether $k$ is positive/negative.
Moreover, if the trajectory passes though point $x_0$ at time $t_0$, one
requires from (\ref{phase}) and (\ref{arbphi}) that
\begin{equation} \label{init}
\left[(E_{n+k} - E_n)t_0 - \int_{x_n}^{x_0} p_n(x) dx
- \int_{x_{n+k}}^{x_0} p_{n+k}(x) dx\right] /\hbar = \pi /2\bmod 2\pi  .
\end{equation}
For the case of a potential energy which is symmetric about $x_0$, this
expression may be simplified via (\ref{bohr}) to give
\begin{equation}
(E_{n+k} - E_{n})t_0 /\hbar = (n+k/2+1)\pi \bmod 2\pi  .
\end{equation}

Channel structures for various potentials have been numerically
observed for a number of examples \cite{stift,kap,loinaz}, and
(\ref{traj}) has
been previously derived in \cite{kap}, via a decomposition of the
probability distribution rather than of $\psi (x,t)$ (which produces an
extra set of trajectories, discarded as ``classical").
Here it is seen that (\ref{traj}) arises directly
from destructive interference of energy amplitudes,
in a manner entirely analogous to the case
of the particle in a one-dimensional box.

It is hoped to further investigate conditions (\ref{traj}) and
(\ref{init}) elsewhere. Here a simple prediction generated by
(\ref{traj}) will be pointed out.
In particular, if the energy eigenvalues $E_n$ increase
slowly over ranges of length $|k|$ (at least for values of $n$ for which
the amplitudes $A_n(x)$ are significant), and $p_n(x)$ varies slowly
over such ranges, then from (\ref{traj}) one has
$dx_k/dt \approx k (dE_n/dn) /(2p_n(x_k))$.
Thus, if two channels coresponding to two values $k$ and $k'$ intersect
at some point in spacetime, their slopes at the point of
intersection are predicted to be
approximately in the ratio
\begin{equation}
(dx_k/dt)/(dx_{k'}/dt) \approx k/k'  .
\end{equation}
This provides a simple test of the applicability of this approach to a
given quantum carpet structure:  the strongest channels, corresponding
to small values of $|k|$, are predicted to intersect with slopes related
by simple rational numbers.
Note from (\ref{arb}) that conditions (\ref{traj}) and (\ref{init}) for
destructive interference need
in fact only hold over the range of $n$ for which the amplitudes
$A_n(x)$ are
significant, to ensure effective destructive interference.  This is
fortunate, as these conditions cannot in general hold for all $n$;
however, it implies that this approach can in general only be applicable
to
superpositions of a relatively narrow band of energy eigenstates.

\newpage
{\bf FIGURE CAPTIONS}

{\bf Figure 1.}  Density plot of (part of) the probability landscape for
the initial wavefunction in (\ref{fig1}), with $N=20$. 
The dark channel observed to run
from the bottom righthand corner to the top lefthand corner of the plot
corresponds to the trajectory $x=L/3-3Vt$, and is not predicted by the
destructive interference approach.

{\bf Figure 2.}  Plot of $1+S_{-3}(z)$ for the initial wavefunction in
(\ref{fig1}), with $N=20$.  From (\ref{aver}), the sharp minimum
in the vicinity of $z=1/3$ corresponds to a channel in the probability
landscape, along the
trajectory $x=L/3-3Vt$ (as observed in Figure 1).

{\bf Figure 3.}  Approximate reconstruction of the density plot in
Figure 1 via the travelling wave decomposition (\ref{trav}), where only
terms with $|k|\leq 5$ have been included.  All linear structures of
slope less than or equal to 5V in magnitude are successfully reproduced.

{\bf Figure 4.}  The structure functions $S_2(z)$ (solid line), $S_4(z)$
 (dot-dashed line) and $S_6(z)$ (dotted line)
for the case of a uniform initial wavefunction, plotted via
(\ref{unif}).

{\bf Figure 5.} The structure function $S_1(z)$ for the approximate
Gaussian initial wavefunction in (\ref{gauss}) with $\overline{x}=L/3$
and $\sigma=L/40$,
for the cases $\overline{p}=0$ (dashed line) and
$\overline{p}=15\pi\hbar/L$ (solid line).

{\bf Figure 6.}  Density plot of the probability landscape for the
approximate Gaussian initial wavefunction in (\ref{gauss}) with
$\overline{x}=L/3$, $\sigma=L/40$ and 
$\overline{p}=15\pi\hbar/L$.  The detailed shape
of the linear structure running from the bottom lefthand corner to the
top righthand corner corresponds to that of the structure function of
Figure 5 (solid line) in the neighbourhood of $z=0$. 
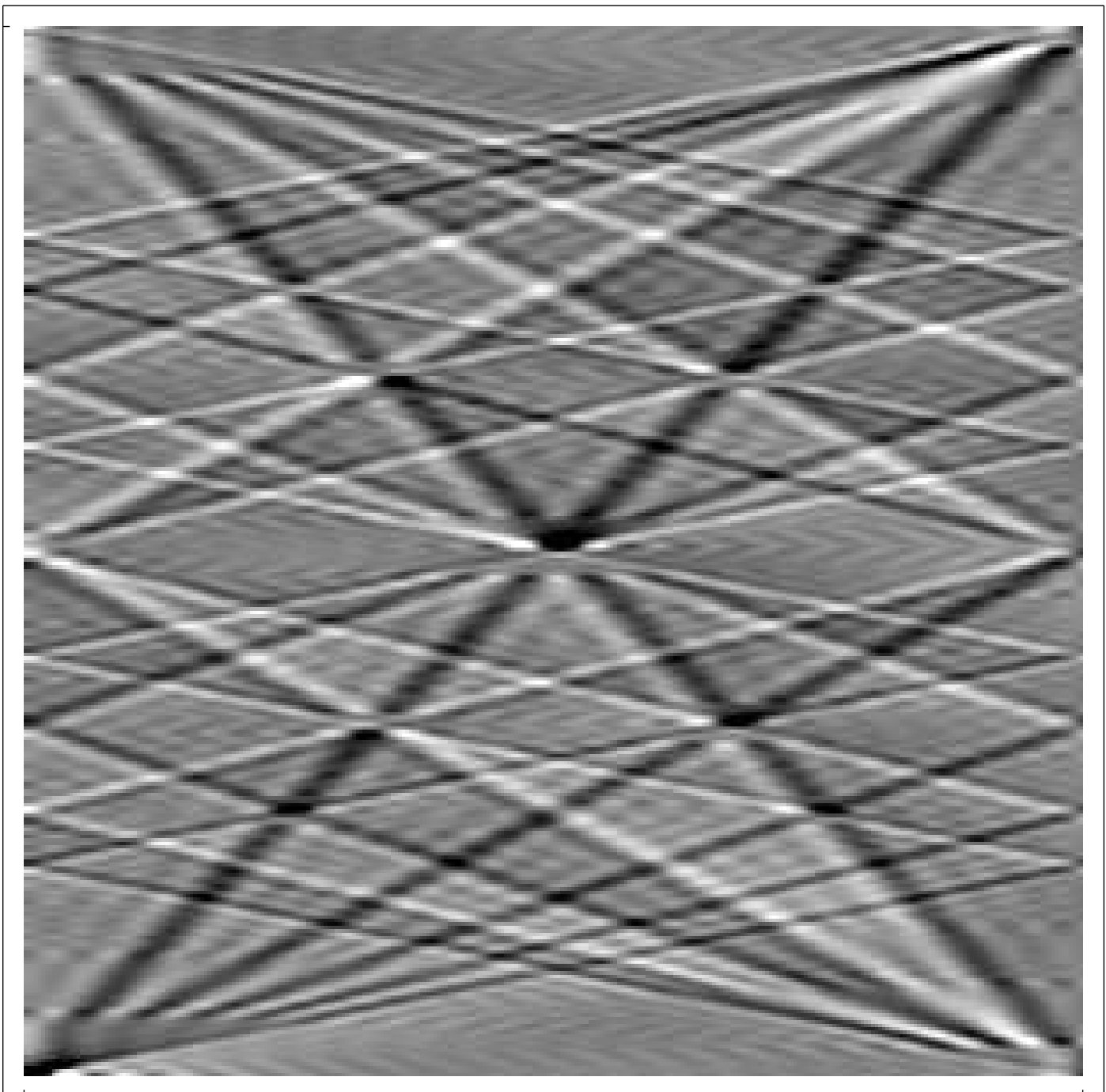
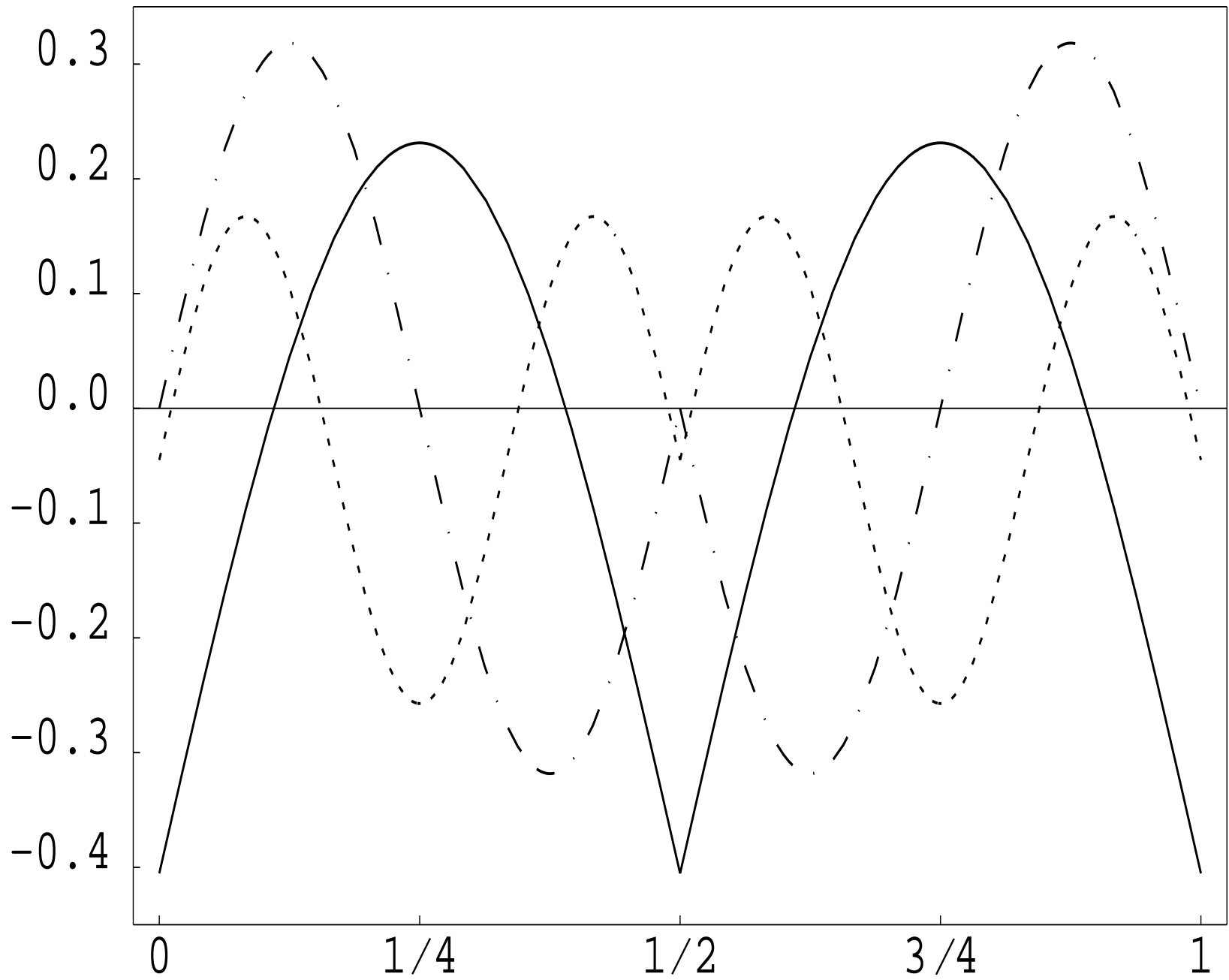
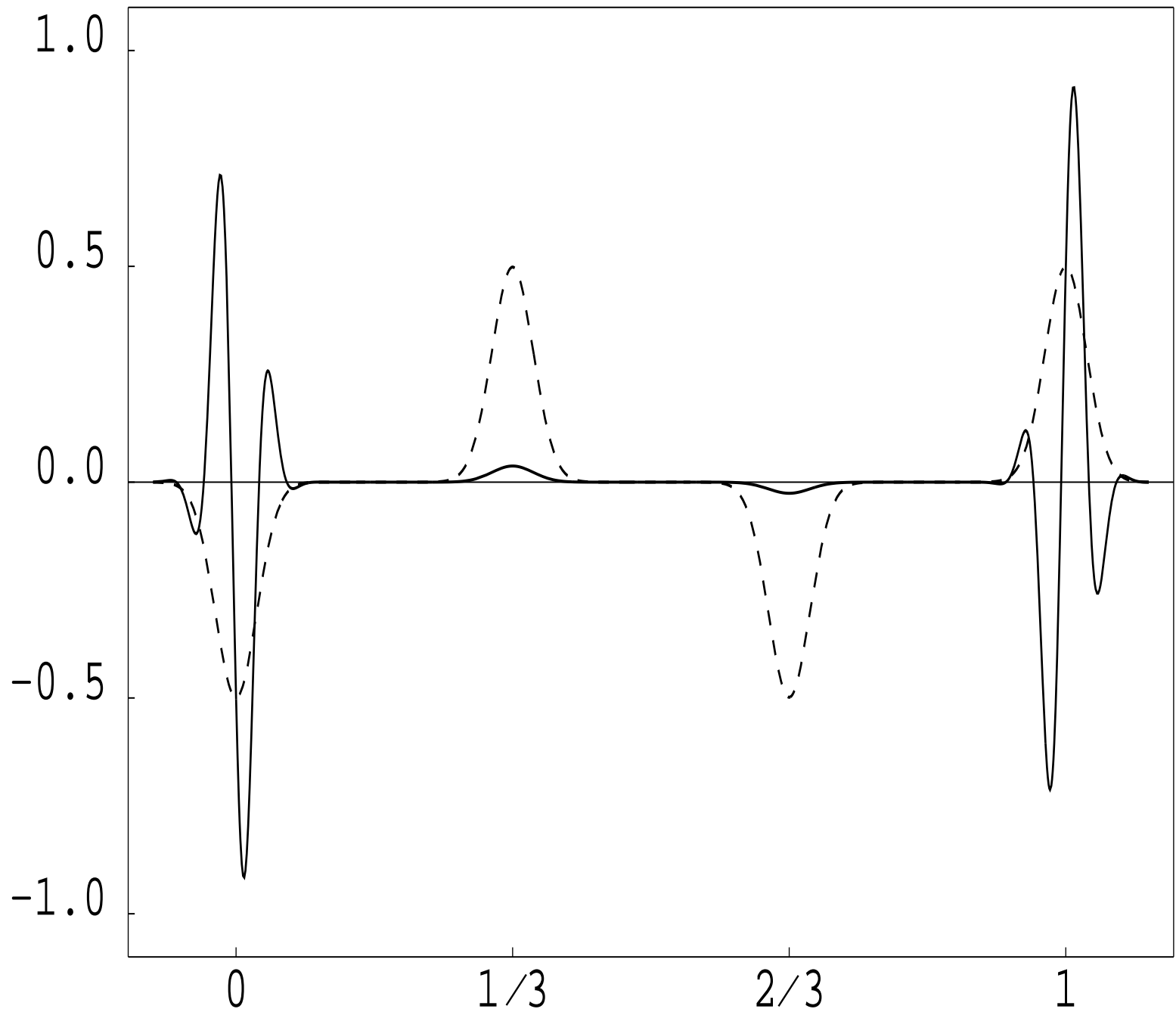
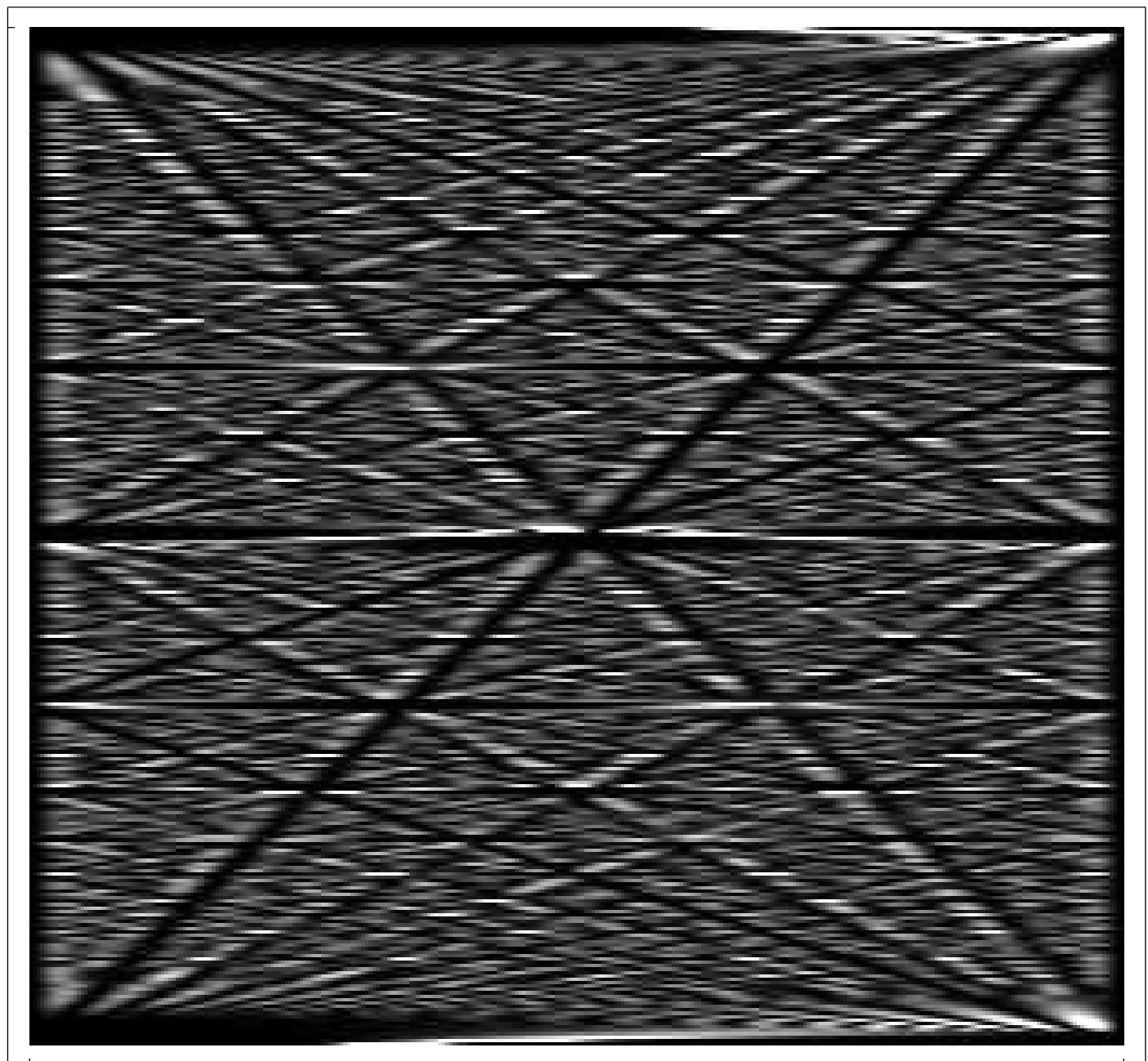
\end{document}